**Quadruply Bonded $Mo_2$ Molecules Acting as an Inborn Emitter-Resonator Quantum System in Free Space**

Miao Meng, Ying Ning Tan, Zi Cong He, Zi Hao Zhong, Jia Zhou, Yu Li Zhou, Guang Yuan Zhu, Chun Y. Liu*

Department of Chemistry, College of Chemistry and Materials Science, Jinan University, 601 Huang–Pu Avenue West, Guangzhou 510632, China

* Correspondence: tcyliu@jnu.edu.cn

*Dedicated to the discovery of metal–metal quadruple bonds by F. Albert Cotton 60 years ago*

**Abstract**

In recent decades, significant progress has been made in studying individual quantum systems composed of atoms, molecules, and photons, with great potential for quantum information processing and other emerging technologies. Here, we show that the quadruply bonded $Mo_2$ unit of a complex molecule can trap visible light between two Mo atoms under ambient conditions, creating a quantized local electromagnetic (EM) field. Our results indicate that both the electronic and vibrational states of the $Mo_2$ molecule are modified by coherent coupling with the scattering field of the $Mo_2$ unit, as evidenced by the Rabi doublet and the Mollow triplet in resonance fluorescence and Raman spectra. The $Mo_2$ molecule allows optical experiments to be carried out in free space, enabling the observation of quantum phenomena using conventional spectroscopic instrumentation. The knowledge gained from this study advances our understanding of metal-metal bond chemistry and quantum optics, with implications for quantum information processing, optoelectronic devices, and chemical reactivity control.

## Introduction

Coherent coupling between atoms and photons is generally achieved in a confined structure containing one of the two components, known as a "box",[1] and then the other is introduced under precise control of the optical physical parameters, including the number of atoms and photons. Quantum optical experiments have long faced significant challenges in both instrumentation and operational techniques. For instance, in the Wineland′s ion trapping experiments, individual ions are confined at a cryogenic temperatures using laser cooling techniques,[1,2,3,4] while in the Haroche′s photon box experiments, [1,5,6] the Rydberg atoms with the main quantum number as high as 50 are prepared to interact with photons in a micromaser. In these controlled environments, where perturbations are minimized, atomic transitions interact coherently with oscillating EM fields, allowing the generation of exciton-polaritons—quasi-Bose-Einstein particles that combine the characters of both atoms and photons. The eigenstates of such hybrid systems are typically described by the Jaynes-Cummings Hamiltonian,[7,8,9] which models the resonant coupling between a two-level atom and a single photonic mode at the single atom-photon level.

In recent decades, significant progress has been made in coupling single molecules with light, which is achieved through narrowing the laser illumination space to focus on a single or a few isolated molecules embedded in a condensed matrix (solvent, crystal, or polymer).[10],[11] This approach creates individual molecular quantum systems characterized by a very small effective photonic mode volume and a moderate Q value, [12] in contrast to the high-Q atomic systems in a micromaser. These differences enable the exploration of light-matter interactions in a new regime, where the intermediate or small Q values and atomic-scale mode volumes provide unique insights into quantum dynamics and transitions, whereas the Purcell effect remains pronounced, as demonstrated in the $Ni_2$ system with $Q \approx 7$.[13] This ability to precisely control light-matter interactions at the molecular scale has provided insights into quantum phenomena such as photon antibunching, [14,15] squeezed light, [16,17] vacuum Rabi splitting[18,19] and Mollow triplet,[11,13,20] in the single-molecule systems. Building on these fundamental advances within the framework of the Jaynes-Cummings formalism,[7] researchers have opened new avenues for probing and manipulating molecular processes in chemistry and biology. These breakthroughs offer unparalleled precision for investigating molecular functions and interactions, with implications for areas such as single-molecule-single-photon spectroscopy,[21] quantum sensor and devices.[16,22] By bridging the fields of quantum optics, chemistry, and biology, single-molecule light-matter coupling represents a unique platform for exploring quantum states and achieving molecular-level control over light.

In order to reduce the photonic mode volume $V$ and thus increase the field amplitude for strong matter-light interactions, various plasmonic cavities with two nanoparticles spaced by a nanometer gap have been intensively studied over the past decades.[23] Using the nanoparticle-on-the-mirror (NPOM) technique, the size of the gold sphere has approached the atomic scale, and the effective mode volume is reduced to $10^{-2}$ $nm^3$.[24] Studies in quantum optics indicate that when the interfacial gap ($d$) of a metallic sphere dimer is further reduced to the contact region ($d \leq 0$), tip-to-tip charge transfer plasmon (CTP) occurs *via* quantum tunneling,[25,26] generating the scattered EM field that is useful for driving a light-matter interaction in the visible region.[23] Recent theoretical and experimental advances have led us to investigate the quantum optical responses of dinuclear transition complex molecules with a $Mo_2$ (ref. 27) or $Ni_2$ (ref. 13) core, where the metal-metal separation falls in the quantum tunneling regime. Spectroscopic results showed that the electronic transitions in the $Mo_2$ complexes are split and shifted, revealing the formation of the exciton-polaritons in cavity-free solution.[27] In-depth study on the $Ni_2$ complex demonstrated that the collective coupling of $N$-molecule ensembles scales as $N\sqrt{N}\Omega$,[13] which is significantly different from the $\sqrt{N}\Omega$ scaling in the Tavis-Cummings model.[28] The $M_2$ (M = Ni, Mo, or other transition metal atoms) quantum system enables control over quantum optical phenomena at the molecular level, achieving field quantization without a cavity.[6] In particular, the quadruply bonded $Mo_2$ unit with the interfacial distance $d < 0$ becomes a desirable testbed for exploring the limit of quantum optical effects,[25] with the potential to construct the practical ultrasmall qubit in quantum circuits.

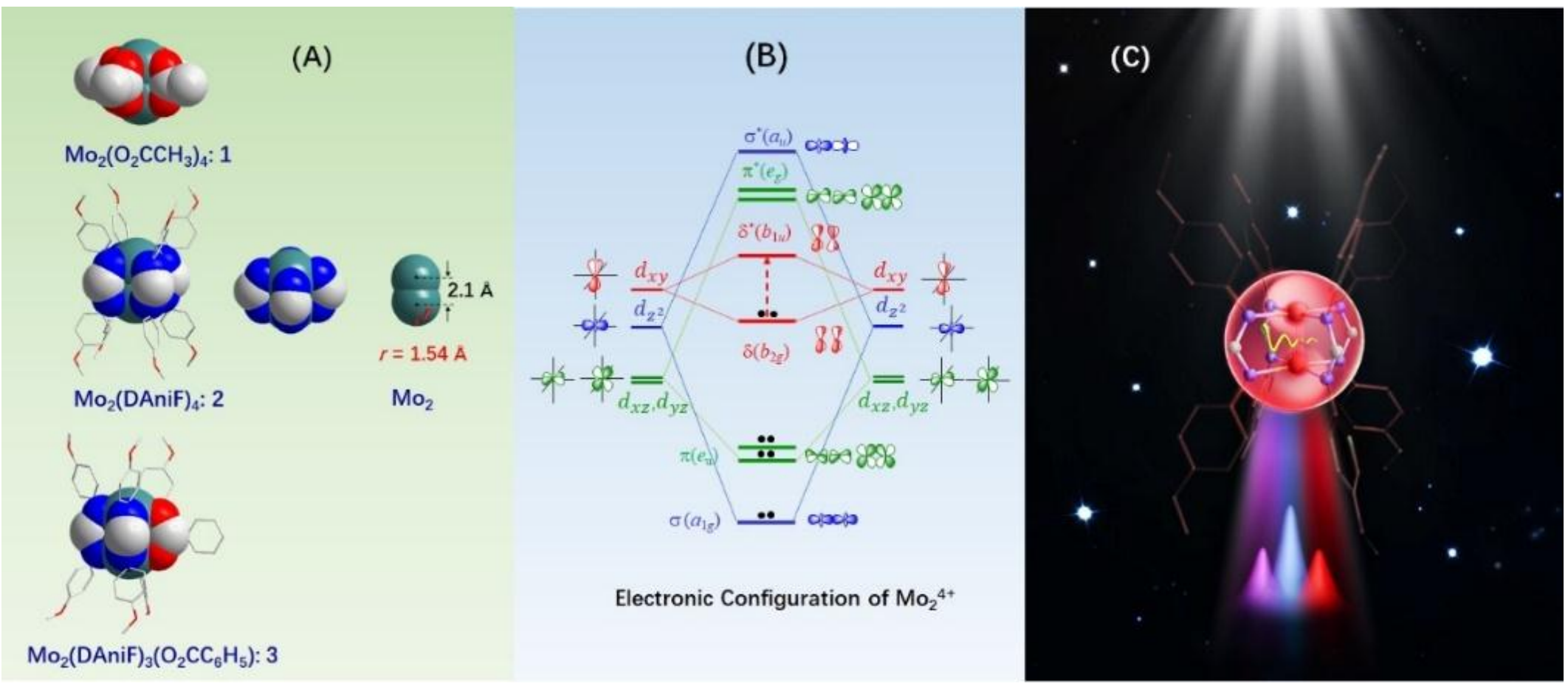

**Figure 1. Molecular structure, electronic configuration, and resonant optical coupling of the $Mo_2$ complexes.** (A) Space-filling models for $Mo_2(O_2CCH_3)4$ (**1**), $Mo_2$(form)$_4$ (**2** and **2′**), and $Mo_2(DAniF)_3(O_2CC_6H_5)$ (**3**). (B) The electronic configuration of $\sigma^2\pi^4\delta^2$ for a quadruply bonded $Mo_2$ unit. The characteristic $\delta\rightarrow\delta*$ transition is indicated by the dotted vertical line with an upward arrow. (C) Illustration of the evolution of the Mollow triplet fluorescence in the $Mo_2$ molecular system (**2**) when illuminated with visible light in free space.

In 1964, a *Science* paper by F. Albert Cotton[29] elucidated the molecular structure of $Re_2Cl_8^{2-}$ in terms of Pauling′s valence bond theory, which recognizes the quadruple bonds between the two rhenium (III) atoms in the molecule and pioneered the dimetal coordination chemistry.[30] The formation of quadruple bonds between the two transition metal ions removes the degeneracy of the atomic orbitals, resulting in an electronic $\sigma^2\pi^4\delta^2$ configuration for the $M_2$ ($M^{n+} = Re^{3+}$, $Cr^{2+}$, $Mo^{2+}$, $W^{2+}$) cores, as shown in Figure 1B. In the $\sigma^2\pi^4\delta^2$ scheme, the $\delta$ orbital is the highest occupied molecular orbital (HOMO) of the $M_2$ unit, and the $\delta$* orbital becomes the lowest unoccupied molecular orbital (LUMO).[30] This energy diagram endows a quadruply bonded dimetal complex with the metal-based close shell ground state and the characteristic electric dipole and spin allowed $\delta$ ($b_{2g}$)$\rightarrow\delta$*($b_{1u}$) (HOMO to LUMO) transition, i.e., the $A_{1g} \rightarrow A_{2u}$ transition for the $M_2$ molecules in the point group $D_{4h}$. Since the discovery of the quadruple bond, metal-metal bonds with a bond order higher than four have attracted considerable interest.[31,32] In particular, quintuple and sextuple bond have been proposed for molybdenum dimer ($Mo_2$) in the gaseous state in matrices, which is supported by the 0.1Å shorter Mo-Mo bond and relatively high bond energy.[31,33] However, the spectral behavior of these "naked" $Mo_2$ species has not been clearly elucidated with respect to the proposed Mo-Mo bond order.[34,35]

In addition, some quadruply bonded $Mo_2$ complexes exhibit unusual spectroscopic properties that cannot be understood from their electronic structures within the $\sigma^2\pi^4\delta^2$ scheme. For example, for $Mo_2(TiPB)_4$, where TiPB = 2,4,6-triisopropylphenyl carboxylate,[36] the characteristic $\delta \rightarrow \delta^*$ transition (Figure 1B), expected to appear at ~ 440 nm, is not visible in the absorption spectrum, whereas the singly oxidized $[Mo_2(TiPB)_4]^+$ shows an intense absorption band at 550 nm. In our study of the photoinduced electron transfer (ET) in the donor ($Mo_2$)-bridge-acceptor ($Mo_2$) complexes,[37] anomalous ET kinetics were obtained from the ultrafast transient spectroscopic study, showing that the donor-acceptor ET is faster than the charge recombination from the bridge to the $Mo_2$ center by an order of magnitude. Furthermore, the phenylene-bridged $Mo_2$ dimers[38] exhibit the intervalence charge transfer (IVCT) bands in the infrared (IR) region, i.e., 4800 -2450 $cm^{-1}$, whereas for the pyrazine-bridged diruthenium mixed-valence complex, known as the Cruetz-Taube ion, the IVCT band is observed at 6410 $cm^{-1}$ in the near-IR region.[39] The exceptionally low energy Franck-Condon barrier for long-range intramolecular electron transfer in the $Mo_2$ dimers suggests an energy gain by coherent coupling of the molecule with light.[27]

In this work, we have investigated four $Mo_2$ complexes in terms of light-matter interaction, including $Mo_2(O_2CCH_3)_4$ (**1**), $Mo_2(DAniF)_4$ (**2**) (DAniF = *N, N′*-di(*p*-anisyl)formamidinate), $Mo_2(DTolF)_4$ (**2′**) (DTolF = *N, N′*-di(*p*-tolyl)formamidinate), and $Mo_2(DAniF)_3(O_2CC_6H_5)$ (**3**) (Figure 1A, see Supplementary Figures 1, 2, 3 and 4). In previous work,[40,41,42] the molecular and electronic structures and UV-visible spectra of **1**, **2**, **2′** and the singly oxidized [**2′**]$^+$ were studied from a pure chemical perspective. Complexes **2** (**2**$^+$) and **3** (**3**$^+$) were included in our primary study of the molecule-photon interaction in the $Mo_2$ molecular systems.[27] These molecules exhibit a paddlewheel structure for the coordination shell with the two Mo atoms separated by 0.21 nm and $d = -0.1$ nm $< 0$ (Figure 1A).[27,41,42] The two $Mo^{2+}$ ions in the $Mo_2^{4+}$ unit are about 0.04 nm apart, falling in the contact regime.[25]

## Results and Discussion

**Quantum Optical Responses of the $Mo_2$ Complexes and Natural Frequency of the $Mo_2$ Resonator.** Complex **1** exhibits a single broad fluorescence band with a maximum at ~ 400 nm when excited by the second-harmonic generation (SHG) of a fundamental laser beam with a wavelength $\lambda_{ex}$ varying between 560 nm and 600 nm (Figure 2A). This peak does not correspond to the decay of the excited state (δ*) of the $Mo_2$ unit to the ground state (δ), which occurs at ~ 435 nm (23000 $cm^{-1}$),[43] nor does it align with any other defined electronic transition. This fluorescence behavior is similar to what has been observed for $Cu_2(O_2CCH(C_2H_5)(CH_2)_3CH_3)_4$ ($Cu_2$)[27] and $Ni_2(DAniF)_4$ ($Ni_2$),[13] both of which have unbound dimetal units and fluoresce at 420 nm and 380 nm, respectively. We attribute this fluorescence to scattering of the dimetal ($M_2$) unit upon excitation, analogous to the CTP modes observed in nanoparticle dimers at the sub-nanometer scale.[23,25] The scattering field around the $Mo_2$ center ($\omega_{Mo2}$) is extremely intense, primarily due to the Purcell effect resulting from the atomistic confinement of the $Mo_2$ unit. [13,23,25] This enhanced local EM field propagates traversing the Mo–Mo bond axis (Figure 2B), which is represented by a characteristic fluorescence spectrum for different $M_2$ nuclearities. The transition energy defines the natural frequency of the $M_2$ unit, functioning as a dimetal resonator.[13] For example, the $Mo_2$ unit, with the shortest M(II)···M(II) distance (0.04 nm), has a natural frequency (400 nm) that falls between $Ni_2$ ($\omega_{Ni2}$ = 380 nm, $d_{Ni2}$ = 0.124 nm) and $Cu_2$ ($\omega_{Cu2}$ = 420 nm, $d_{Cu2}$ = 0.118 nm).[13,27] Of the three $M_2$ units, this natural frequency varies depending on the M(II)–M(II) distance, following a similar observation for nanoparticle dimers in the quantum tunneling regime.[25] This suggests a critical distance $d_{QR}$ ~ 0.12 nm for the $M_2$ systems, which is significantly smaller than $d_{QR} \approx 0.31$ nm observed in atom-scale plasmonic cavities.[25] Thus, this suggests that the fluorescence of the $M_2$ complexes could be a quantum optical effect related to the geometry of the dimetal unit, regardless of the coordination environment influencing the electronic structure. The reduced $d_{QR}$ for $M_2$ units and the blue shift of the natural frequency ($\omega_{M2}$), compared to those for the metal sphere dimers, are attributed to the extremely short M···M distances and the positive charge of the $M^{2+}$ ion.

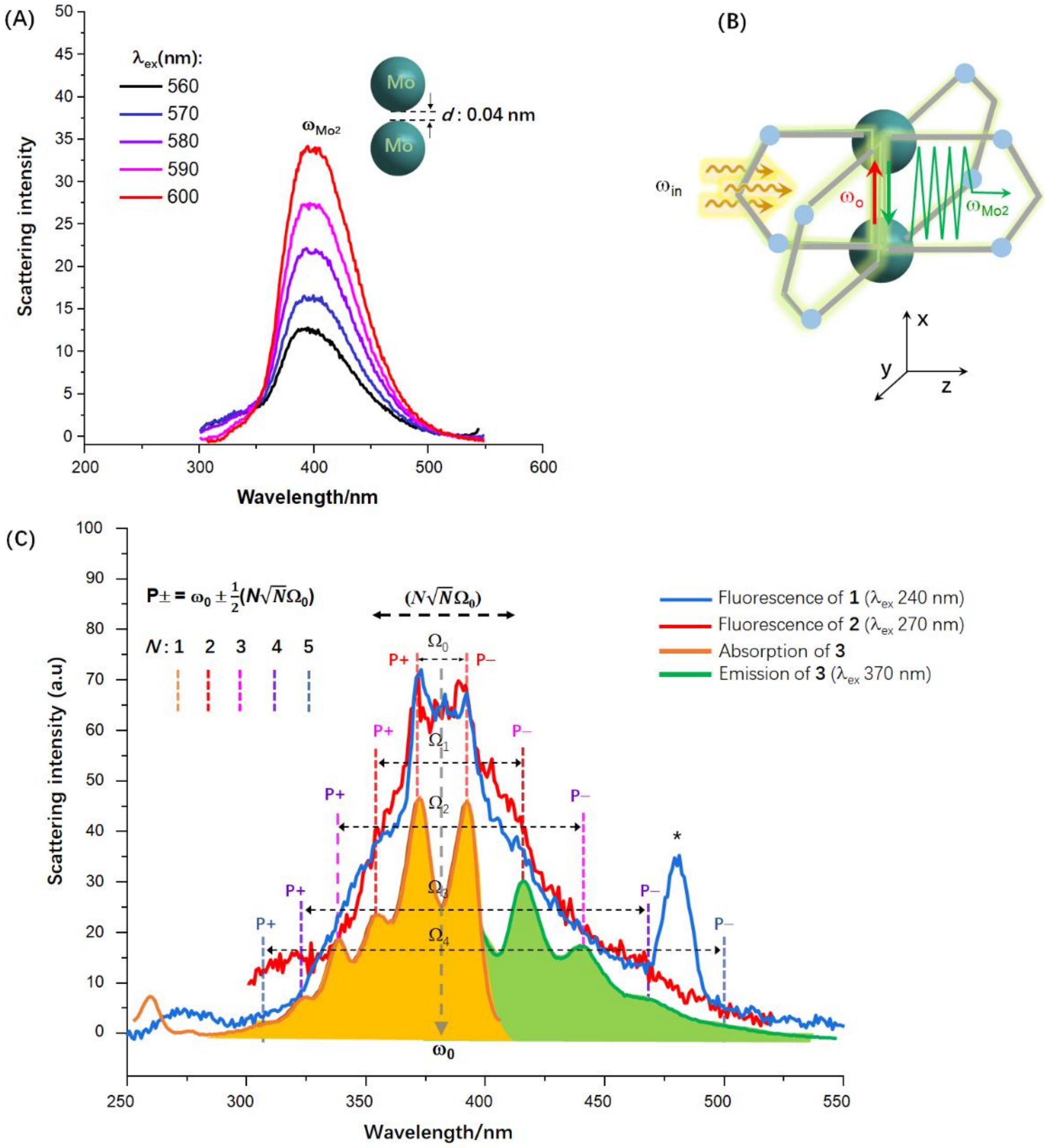

**Figure 2. Optical responses and the interpretation of the $Mo_2$ complex molecules.** (A) Fluorescence of $Mo_2(O_2CCH_3)_4$ (**1**) showing the thermally averaged scattering continuum of the $Mo_2$ unit under incoherent excitation. (B) A pictorial illustration of the conversion of incident classical light ($\omega_{in}$) into quantum light ($\omega_{Mo2}$) through the $Mo_2$ unit by charge transfer between the two Mo atoms. (C) Fluorescence spectra of the $Mo_2$ complexes showing the Rabi doublets of the *N*-molecule ensembles through collective coupling of different orders with *N* up to 5. The asterisk peak indicates the double frequency signal of laser excitation.

Excitation of **1** and **2** with monochromatic beams of $\lambda_{ex}$ = 240-270 nm results in the resonance fluorescence featuring pairs of emission peaks symmetrically distributed around the 380 nm position, as shown in Figure 2C. This unusual fluorescence profile has been observed for the $Ni_2$ analog, $Ni_2(DAniF)_4$, with a weak, narrow central peak corresponding to the resonance energy $\omega_0$ (380 nm),[13] which is the dark state of the $Ni_2$ emitter. In the $Ni_2$ system, this resonance fluorescence obtained by incoherent excitation has been attributed to the vacuum Rabi splitting of the two-level CT mode ($\omega_0$) of $Ni_2$, which is resonantly coupled to the scattering field $\omega_{Ni2}$. Therefore, in the spectra of $Mo_2$ (Figure 2C), the two peaks at 370 nm and 390 nm centered at $\omega_0$ are assigned to the exciton-polaritons P+ and P− for the single molecules, giving a coupling rate ($\Omega_0$) of 1380 (±10) $cm^{-1}$, almost identical to that of $Ni_2$. Additional sidebands are observed symmetrically distributed over the spectrum across the resonance, representing collective coupling of the *N*-molecule ensembles to the scattering field.[13] The intensity distribution of the spectra reflects the statistical distribution of the molecular ensembles in solution, indicating a lower concentration for the higher order (larger *N*) molecular array. The Rabi splitting states for high-order collective coupling,[13] which are not well resolved in the luminescence spectra, are confirmed by the combined excitation-emission spectra for **3** (Figure 2C). The measured collective coupling rates, 4430 $cm^{-1}$ ($N$ = 2), 7140 $cm^{-1}$ ($N$ = 3), 9640 $cm^{-1}$ ($N$ = 4), and 13000 $cm^{-1}$ ($N$ = 5) (see Supplementary Table 1), are similar to those observed for the $Ni_2$ system,[13] confirming the $N\sqrt{N}\Omega_0$ scaling for collective coupling of the *N*-molecule ensemble. In this collective coupling scaling, the factor *N* results from the simultaneous excitation of the *N* molecules (spins), while the factor $\sqrt{N}$ is due to the absorption of the *N* photons of the ensemble for the collective excitation or the *N* photons contributed by the *N* $M_2$ resonators with one for each, which are resonantly and individually coupled to the emitters.[13] It differs significantly from that in the Tavis-Cummings model,[28] where the $\sqrt{N}$ factor represents the suppression of the full excitation of the Dicke states by the strong dipole-dipole interaction,[1] indicating the dipole blockade in the two-atom Dicke model.[13,44,45] It is remarkable that this collective coupling scaling is *N* times larger than $\sqrt{N}\Omega_0$ in the Tavis-Cummings model by breakdown of the dipole blockade,[28] allowing ultrastrong coupling to be easily achieved.

The parametric agreement in the resonance fluorescence between the $Mo_2$ and $Ni_2$ systems suggests a common physical origin for the optical events in both the bonded and unbonded $M_2$ systems. It is well known that the electronic structures and electronic spectra for the unbonded $Ni_2(DAniF)_4$ (ref.13) and the quadruply bonded **1**[40,43] and **2** (**2′**)[41,42] with different coordination environments are distinct from each other. This rules out the possibility that this featured fluorescence for these complexes results from a two-level electronic excitation or a Frenkel exciton of the molecule (emitter) coupled to the scattering field. Instead, it is the M↔M CT mode $\omega_0$ that is coupled to the scattering field ($\omega_{M2}$).[8,13,23] Importantly, the very similar fluorescence spectra for these two complexes suggest that the $Ni_2$ and $Mo_2$ units share very similar CT resonance energies, despite their substantial geometric and coordinative differences. It appears that $\omega_0$ = 26300 cm$^{-1}$ (380 nm) represents the quantum limit for the $M_2$ units in the conductive contact regime ($d_{QR} \geq d \leq 0$ nm).

**Transition from the Rabi Splitting States to the Mollow Triplet States.** Complex **1** exhibits distinct fluorescence profiles with a gradual decrease of the excitation energy to $\lambda_{ex}$ = 400 nm, as shown in Figures 3A and 3B. In the fluorescence spectrum at $\lambda_{ex}$ = 240 nm, two pairs of peaks are observed symmetrically spread over a sharp $\omega_0$ peak at 382 nm, which are attributed to the Rabi doublets for the single molecules and the two-molecule ensemble (Figures 2C and 3A, blue). With 250 nm excitation, the fluorescence spectrum shows an asymmetric triplet at 372 nm, a position corresponding to the P+ peak for the two-molecule ensemble,[13] as shown in Figure 3A. Remarkably, this feature is duplicated at 745 nm, half of the transition energy, accomanied by two satellites at 708 nm and 786 nm, respectively, characterizing a Mollow triplet with half the zeroth oder coupling strength ($\Omega_0/2$), i.e., 700 cm$^{-1}$. This fluorescence at half of the energy scale, resulting from optical parametric generation (OPG) of the resonance fluorescence, demonstrates excitation of the $\omega_{372}$ sideband by the scattering field,[46] generating the Mollow triplet at $\omega_{372}$ with $\Omega_0' \approx$ 1400 cm$^{-1}$, which verifies the vacuum Rabi splitting for the single molecules. Similar results are observed in the $Ni_2$ system, where the Rabi peaks for the single molecules are duplicated at half of the coupling energy scale, producing the two Rabi components separated by ~700 cm$^{-1}$, exactly half of the zeroth order coupling strength.[13] However, these Mollow triplet profiles are distorted by incoherent excitation and possibly by the squeezed vacuum field of the system at the single-molecule level.[13,16,47,48] According to a recent theoretical study in the strong coupling regime,[49] frequency shifts of the physical modes are correlated with the squeezing parameters of virtual excitations. The emission features at half the coupling scale could signify virtual photon emission from the ground state, representing transitions enabled by the squeezed scattering field. Nevertheless, these observations underscore the intrinsic consistency between the quantum nature of the system and its optical behavior as a JC molecule.[13,47,48]

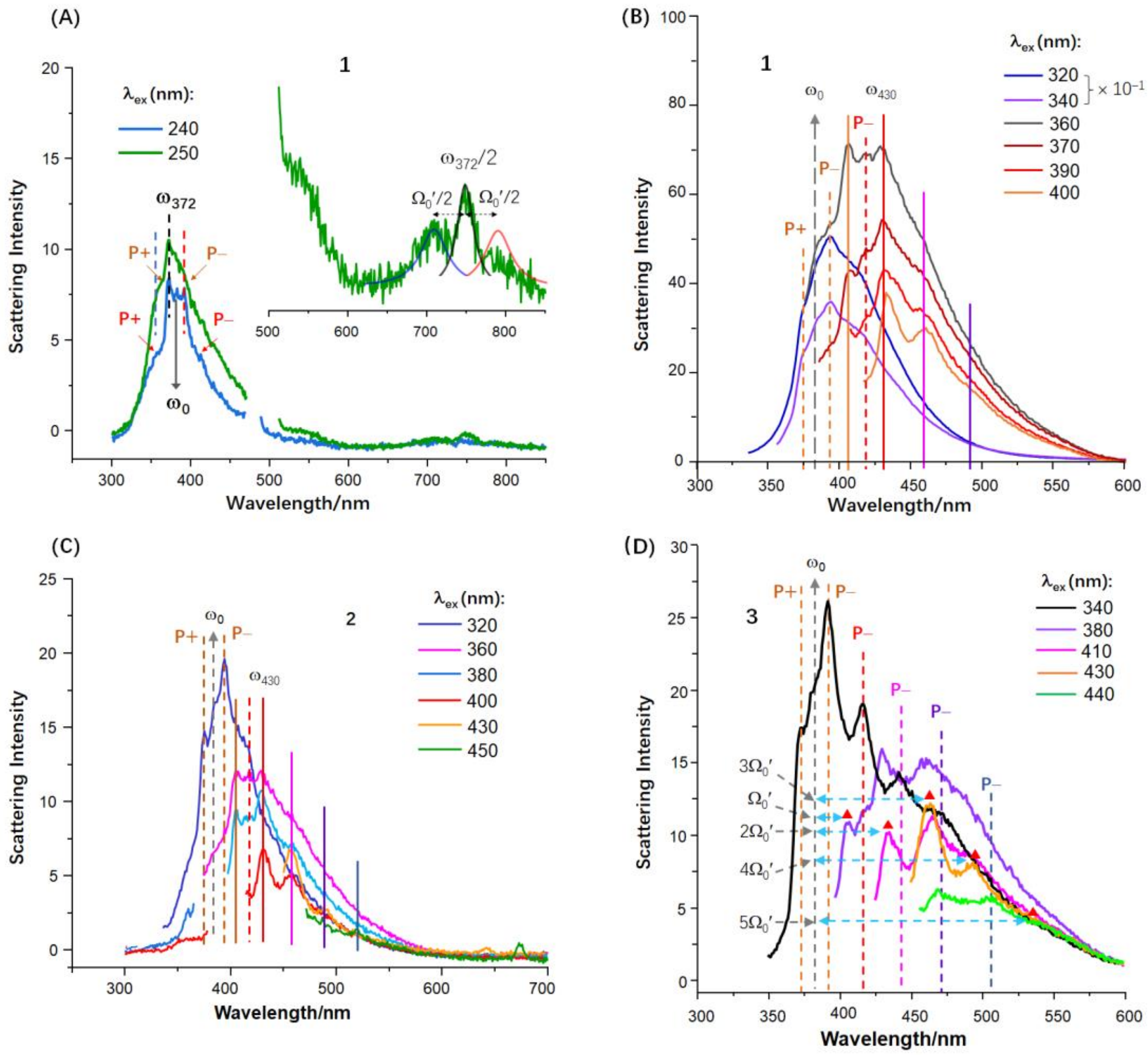

**Figure 3. Quantization of classical visible light by the $Mo_2$ unit.** (A) Resonance fluorescence of **1** with incoherent excitation. The spectrum excited at 240 nm shows symmetrically

distributed peaks around $\omega_0$, indicating the coupling strength $\Omega_0$ for the single molecules, whereas excitation at 250 nm produces an asymmetric Mollow triplet at 372 nm, a position corresponding to the P+ band of $\omega_0$. This feature is duplicated at half of the P+ frequency, i.e., 744 nm, giving rise to half of the coupling strength, i.e., $\Omega_0/2$. (B) Spectral variations of **1** with varying excitations, showing the Mollow triplets derived by exciting the sidebands and their transformation across the Rabi doublet at 417 nm. The colored vertical solid and dashed lines mark the Rabi splitting and the Mollow triplet states, respectively. (C) Featured fluorescence spectra for **2**, showing the ladder structure of the dressed states with progressively lowering the excitation energy. (D) Featured fluorescence spectra for **3**, showing the transformation of the Rabi states into the Mollow states with excitations of $\lambda_{ex}$ > 380 nm.

With a constant decrease of the excitation energy towards $\omega_{Mo2}$ (~ 400 nm), the emission spectra exhibit alternatively Rabi doublet and Mollow triplet peaks for the collective coupling of different orders, as shown in Figure 3B, with the transition energies confirming the state transitions observed across the $M_2$ (M = Ni and Mo) systems.[13] Notably, while similar transitions have been observed in the $Ni_2$ system,[13] the $Mo_2$ complex shows important differences in the resonant coupling dynamics. Facing the $\omega_{372}$ triplet due to the P+ sideband excitation, an intense triplet structure at 392 nm ($\omega_{392}$) is observed upon excitation at 320 nm and 340 nm (Figure 3B), indicating a Mollow triplet arising from the P− sideband excitation.[46] Note that these triplets differ from the standard Mollow triplet[20] in the relative intensities of the three peaks. This distortion of the spectral shape highlights the field-driven state transitions governed by photon antibunching and the squeezed states.[13,50,51,52] The intense $\omega_{392}$ triplet for the $Mo_2$ system,[27] which was not observed for the $Ni_2$ system,[13] can be explained by the enhanced excitation-photon interaction due to the better energy matching between the natural frequency ($\omega_{Mo2}$) and the Rabi sideband ($\omega_{392}$). When the excitation wavelength exceeds 370 nm, a new triplet appears with a central peak at 430 nm, corresponding to the second low-energy sideband of the Mollow triplet at resonance, i.e., $\omega_{430} = (\omega_0 - 2\Omega_0)$. It is worth noting that the Mollow triplet at $\omega_0$ is not observed for this $Mo_2$ system, whereas in the $Ni_2$ system, the resonance Mollow triplet is extremely weak.[13] These results support that the coupling system is driven by photon antibunching, which is diminished at the resonance for the two-molecule systems. [14,51]

For **2**, the fluorescence at $\lambda_{ex}$ = 320 nm shows the characteristic triplet with a sharp central peak at $\omega_{392}$. Both spectra of **1** (Figure 3B) and **2** (Figure 3C) at 360 nm excitation show a transition of the Mollow triplet from $\omega_{392}$ to $\omega_{430}$ through the Rabi splitting states centered at 417 nm, similar to the transition of the Mollow triplet from $\omega_{372}$ to $\omega_{392}$ across $\omega_0$ (Figure 3A). The $\omega_{417}$ resonance corresponds to the P− Rabi line for the first-order collective coupling (Figure 2C) as well as to the red sideband of the $\omega_{392}$ triplet (Figures 3B and 3C). It is worth noting that for both **1** and **2**, the 417 nm peak in the transition spectrum is narrow and weaker than its Rabi components, indicating that this $\omega_{417}$ sideband is excited by photon antibunching of the Mollow triplet,[46] and consequently, becomes "dark". These spectra clearly show that the energy interval between two adjacent peaks, with one from a Rabi transition and the other from a Mollow transition or vice versa, is ~700 $cm^{-1}$, i.e., $\Omega_0/2$, corresponding to a phase shift of $\pi$, while two neighboring Rabi or Mollow peaks are separated by ~1400 $cm^{-1}$ , i.e., $\Omega_0$ (or $2\pi$). These results indicate that the photon antibunching of the Mollow triplet sidebands enables excitation of the polaritons, which has been theoretically demonstrated recently.[46] Therefore, this transition between the Rabi doublet and the Mollow triplet states implies a phase-sensitive population decay, a characteristic observed across the $M_2$ systems.[13] These results implicate the influence of squeezed scattering fields in breaking down the dipole blockade in the two-atom Dicke model and allowing simultaneous excitation of the *N* molecules in the ensemble.[13,44,45,50,52,53] The fluorescence spectra of **2** show that field quantization by $Mo_2$ provides the lowest energy quantum light with a wavelength of 522 nm when excited at 450 nm (Figure 3C), confirming the results obtained for the $Ni_2$ system.[13]

For complex **3**, excitation at 340 nm allows observation of the low-energy half of the full Rabi splitting spectrum of collective coupling, which exhibits five P− branches for the *N*-molecule ensembles ($N = 1 - 5$) (Figure 3D). The energies of these peaks are predicted by $\omega_0 - (N\sqrt{N}\,\Omega_0)/2$, which satisfies the Rabi splitting $N\sqrt{N}\,\Omega_0$ of the *N*- molecule ensemble.[13] This Rabi spectral profile is transformed into the Mollow multiplet structure at 430 nm by the near-resonance excitation (380 nm). By gradually decreasing the excitation energy, five peaks are probed at 404 nm, 430 nm, 460 nm, 492 nm and 530 nm with an energy interval of 1450 (± 50 ) $cm^{-1}$, as marked by red triangles in Figure 3D. These peaks correspond to the valleys in the spectrum excited at 340 nm, indicating a phase difference of π between the Rabi and Mollow states.[13] These peaks are considered as the red sidebands of a sequence of Mollow triplets, and the transition energies are scaled by $(\omega_0 - N\Omega_0')$ ($N = 1, 2, 3, 4$ and 5) with $\Omega_0' = 1450$ $cm^{-1}$. Thus, the multiple Mollow triplets result from the collective coupling of the *N*-molecule ensembles in solution and thus, share a common physical origin for the multiplicity of the Rabi doublets. The full spectrum represents the quantization of the scattering field of the $Mo_2$ resonator, in excellent agreement with the results obtained in the $Ni_2$ system.[13]

**Vacuum Rabi Splitting of the Electronic Transitions in the Cationic Complexes.** Under the $D_{4h}$ symmetry, the theoretical studies of **1** (ref.40) and the computational model of **2** and **2′**, $Mo_2[(NH)_2CH]_4$,[41] produced essentially the same molecular orbital (MO) diagrams, revealing similar electronic transitions related to the $Mo_2$ center. Experimentally, three bands around 320 nm, 390 nm and 440 nm, designated I, II and III, respectively, are found for each of these $Mo_2$ complexes, and are of particular interest.[40,41,42,43] After some controversy in the assignment of these absorptions in earlier studies,[40,43,54,55] it is now generally accepted that the bands I and II are assigned to the $1a_u$ (O or N) → $5e_g$ (π*) and $4e_g$ (O or N) → $2b_{1u}$ (δ*) transitions,[41,42] respectively, while the lowest energy absorption is the well-defined $\delta(2b_{2g}) \rightarrow \delta^*(2b_{1u})$ transition or III.[41,42,43,55] These transitions are all dipole-allowed. Transitions I and II involve the atomic orbitals of the ligand donors (O or N) in the ground state and the $Mo_2$-based antibonding orbitals π* or δ* in the excited state (Figure 1B), thus belonging to the ligand to metal charge transfer (LMCT) transitions (Figure 4A). In the spectra for the neutral complexes in solution (Figures 4B-4E), the transitions are weak and overlapping, especially for III, which prevents an accurate determination of the transition energies.

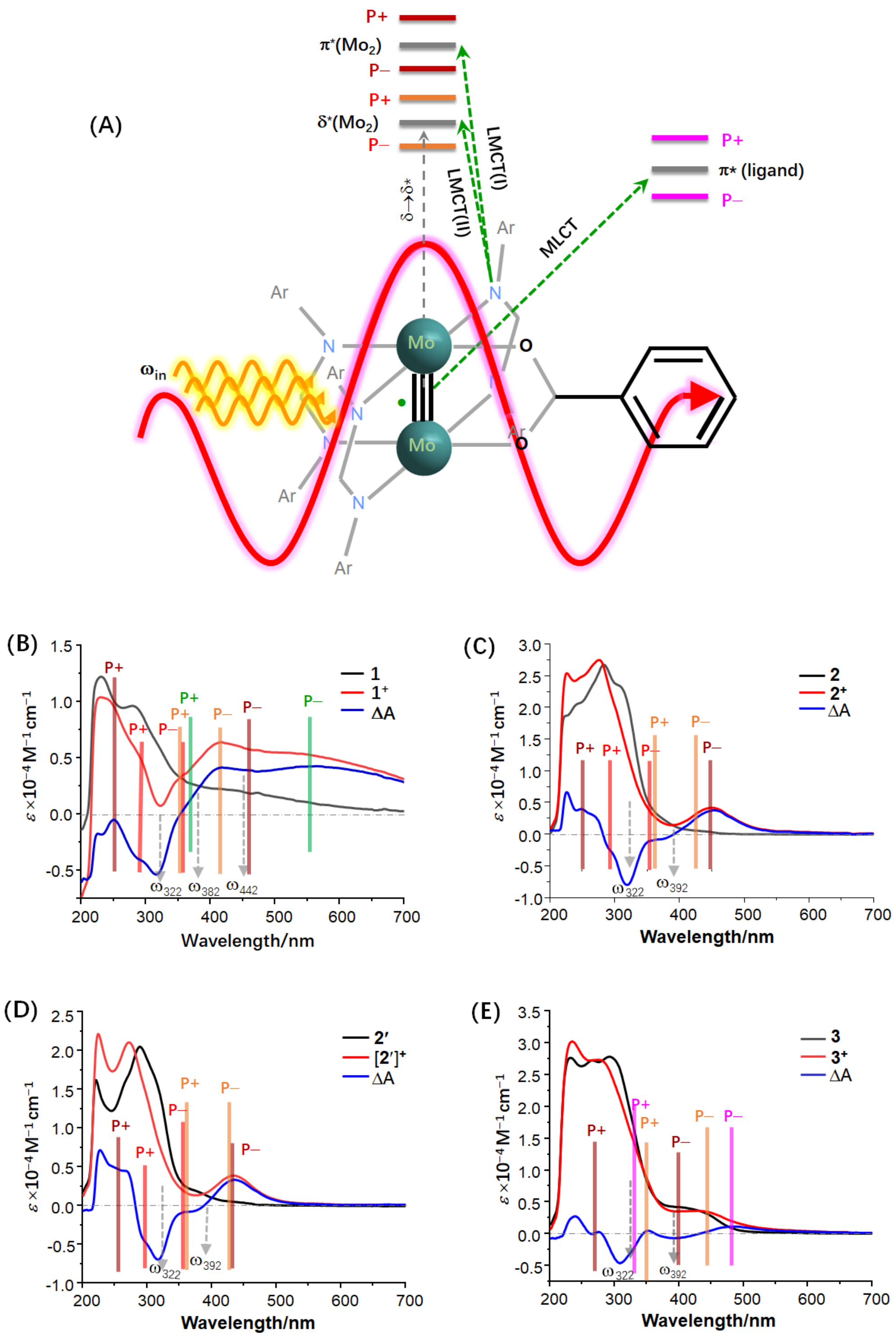

**Figure 4. Vacuum Rabi splitting of the electronic transitions in UV-Vis absorption spectra for the singly oxidized $Mo_2$ complexes.** (A) Resonant coupling of the two-level transitions with the quantized local scattering field in the singly oxidized $Mo_2$ formamidinates. (B), (C), (D) and (E) The UV-Vis. spectra for **1** (**1**$^+$), **2** (**2**$^+$), **2′** ([**2′**]$^+$) and **3** (**3**$^+$), respectively, and the corresponding ΔA spectra. The gray vertical dashed lines with a down arrow indicate the

selected field oscillating modes ($\omega_L$) resonantly coupled to the excitation normal mode. The Rabi bands P+ and P− of different excitations are marked by colored vertical lines: red (dark red) for excitation I, orange for excitation II, green for excitation III, and pink for the MLCT excitation.

The singly oxidized cationic species show intense absorptions that differ significantly from the spectra of the neutral precursors (Figures 4B-4E). This cannot be explained by the electronic structure alone, because theoretical calculations of **2** and **2**$^+$ showed similar MO diagrams with comparable transition energies for the neutral and oxidized molecules.[41] For each complex, the spectral differences between the neutral and cationic complexes are manifested in the different spectrum (ΔA),[27] which is derived by subtracting the molar extinction coefficients of the neutral complex from those of the oxidized complex over the spectral region (Figures 4B-4E). In general, the cationic complex shows more intense and more absorptions compared to the neutral complex,[27] which accounts for the dark color of the compounds.[41] In the ΔA spectra, all complexes show a pronounced spectral valley at ~ 320 nm, corresponding to transition I in energy. For the $Mo_2$ formamidinates **2**, **2′** and **3**, the negative absorptions at ~ 390 nm in the ΔA spectra coincide with band II. The disappearance of the characteristic bands I and II and the appearance of new intense absorptions appearing around the resonances are reminiscent of the optical coupling of the two-level excitations occurring in the cationic species.[27] This optical coupling switches the normal excitation mode to "dark", developing a composite system with the two-level molecule resonantly coupled to a single populated mode of the scattering field,[13, 27] as predicted in the earlier theoretical study.[56] In this context, we interpret the absorption spectra of the positively charged molecules in terms of vacuum Rabi splitting of the electronic transitions.[18]

Discrete quantum lights of wavelengths 322 nm, 382nm, 392 nm and 442 nm are found in the spectrum of the quantized scattering field (Figure 2C), corresponding to $\omega_0$ (382 nm) and the Rabi components P+ ($N$ = 4) and P− ($N$ = 1 and 3) according to the scaling $P\pm = \omega_0 \pm \frac{1}{2}(N\sqrt{N}\ \Omega_0)$. These photonic modes are in remarkable agreement with the resonaces I, II and III for the neutral $Mo_2$ complexes. For the cationic species, the electronic transitions are resonantly or near-resonantly coupled to the vacuum field of the single modes, evolving the correspoding exciton polaritons. In the spectrum of $\mathbf{1}^+$ and the corresponding ΔA spectrum (Figure 4B), two pairs of absorption bands are symmetrically distributed around the position of $\omega_{322}$. The inner pair consists of two bands at 292 nm (P+) and 358 nm (P−), separated by 6314 $cm^{-1}$. The 250 nm peak in the ΔA spectrum marks the P+ component for the second pair, with the P− band located at 452 nm, taking into account the resonant coupling of the 320 nm transition (I) with the $\omega_{322}$ mode and the intense absorption in the expected region. Thus, the outer pair has the P+ and P− branches separated by 17876 $cm^{-1}$ (2.2 eV). Taking the separation of 6314 $cm^{-1}$ for the first pair of Rabi branches as the coupling strength of the single molecules (Ω), the measured coupling strength of 17876 $cm^{-1}$ from the second pair is in remarkable agreement with the collective coupling strength for the $N$ = 2 ensemble when scaled as $N\sqrt{N}\Omega$,[13] i.e., $2\sqrt{2}\Omega$ = 17860 $cm^{-1}$. The primary and the collective coupling of excitation I to $\omega_{322}$ are confirmed by systems $\mathbf{2}^+$ (Figure 4C) and $[\mathbf{2'}]^+$ (Figure 4D) with similar ΔA profiles and similar coupling strengths (see Supplementary Table 2).

The resonant coupling of excitation II with the $\omega_{392}$ light mode is indicated by the shallow spectral valleys at 392 nm for the three $Mo_2$ formamidinate complexes. For $\mathbf{2}^+$ and $[\mathbf{2'}]^+$, the P+ is located at ~363 nm, overlapping with the first P− of I and forming the 350-365 nm absorption plateau in the ΔA spectra (Figures 4C and 4D). For both, the P− of II is located at ~ 427 nm, closely overlapping with the second P− of I and giving rise to the pronounced asymmetric absorption at ~ 450 nm that is absent for the neutral molecules. The relatively low energy of the second P− band of I for $\mathbf{2}^+$, compared to that for $[\mathbf{2'}]^+$, is consistent with the lower transition energy of I for **2** due to the stronger electron-donating of the anisyl groups of the DaniF ligands.[42] This assignment of the 450 nm absorption to the polaritonic transitions is justified by the increased transition dipole moment (Δμ) of the excitation for the LMCT transitions (I and II) for the cationic complex, which enhance the exciton-photon interaction. In contrast, the assignments to the electronic transitions in the traditional chemical sense (e.g., N ($4e_g$) →δ* ($2b_{1u}$)[41] do not explain the large shifting of the absorptions and the increased intensity for the cationic species because the lower $f_{os}$ for the oxidized molecule would decrease the absorption intensity. In addition, the neutral molecule **2′** shows this II band at 375 nm as a faint shoulder,[42] distanced from the 434 nm band for [**2′**]$^+$ by 3625 cm$^{-1}$. This large red shift of the absorption can be perfectly explained by the vacuum Rabi splitting of the resonance, but not by the electronic transition.[41] The transition II for **1**, which cannot be identified in the spectrum (Figure 4B), is expected to have a higher transition energy than those for the formamidinate analogs due to the substitution of the O donors for the N donors. Indeed, the O → δ* transition energy was calculated to be 3.71 eV or 334 nm.[40] Consequently, for **1**$^+$, this excitation is coupled to the higher energy mode $\omega_{382}$, rather than $\omega_{392}$, as evidenced by the absence of the 392 nm valley in the ΔA spectrum (Figure 4B). Resonant coupling of II with $\omega_{382}$ in **1**$^+$ is further supported by the pronounced P− band at 416 nm and the notable shoulder band at 353 nm for the P+ in the spectrum, giving a coupling rate of 4300 cm$^{-1}$, similar to that of **2**$^+$ and [**2′**]$^+$ (see Supplementary Table 3).

The spectrum of the neutral **3** (Figure 4E) shows a broad absorption from 380 to 480 nm that involves three transitions: II (~ 390 nm), MLCT (395 nm), and III (446 nm).[27] For **3**$^+$, the absorption extends to 500 nm upon single oxidation. This extension is likely due to the additional two-level excitation (Figure 4A), i.e., the MLCT that occurs at 395 nm. Optical coupling is the only way to observe a low energy signal for the high energy excitation. Therefore, the P− band of the MLCT transition can be unambiguously located at 483 nm (Figure 4E), which has been probed at the same wavelength by the femtosecond transient and fluorescence spectra,[27] and thus, the P+ branch is located at 330 nm from the resonant coupling to the $\omega_{392}$ mode. The presence of the 330 nm polariton is evidenced by the blue shift and reduction of the I valley in the ΔA spectrum compared to the ΔA spectra for **2** and **2′**. The coupling strength for the MLCT excitation is determined to be 9600 cm$^{-1}$ (1.2 eV) and the normalized coupling rate $\eta$ = 0.19 in the ultrastrong regime. This ultrastrong coupling is obviously due to the large MLCT transition dipole moment (Figure 4A). The broad negative absorption at 392 nm in the ΔA spectrum is attributed to the bleaching of the II and MLCT excitations, as a result of the resonant coupling of these two electronic normal modes with the photonic $\omega_{392}$ mode. For excitation II, the polariton components are located at 350 nm (P+) and 445 nm (P−) (Figure 4E). This P− transition contributes significantly to the dominant absorption for the cationic species, as observed for the other three systems. In the 320 nm region, the ΔA spectrum shows the familiar funnel-shaped profile, indicating the resonant coupling between excitation I and the field mode $\omega_{322}$; however, for $\mathbf{3}^+$, only one pair of Rabi transitions is probed at 270 nm (P+) and ~ 400 nm (P−), and the coupling is less pronounced.

It is interesting to note that the three oxidized formamidinato $Mo_2$ complexes $\mathbf{2}^+$, $[\mathbf{2'}]^+$ and $\mathbf{3}^+$ do not show absorptions that can be attributed to the $\delta \rightarrow \delta^*$ electronic transition (Figure 1B) and that result from vacuum Rabi splitting of the excitation (III). The ΔA spectra do not show a spectral valley around the 440 nm position corresponding to the vertical transition (III) for the neutral molecule. These results indicate that the $\delta \rightarrow \delta^*$ transition is not coupled to the scattering field of the $Mo_2$ unit. In contrast, the oxidized $\mathbf{1}^+$ shows a broad absorption ranging from 500 to 650 nm with a maximum at 550 nm and a shallow spectral valley at ~450 nm in the ΔA spectrum (Figure 4B), signaling the coherent coupling of III with light. In the study of $Mo_2(TiPB)_4$,[36] no $\delta \rightarrow \delta^*$ transition is observed for the neutral complex, but an intense absorption at 550 nm was observed in the spectrum for the oxidized species $[Mo_2(TiPB)_4]^+$, accompanied by a high-energy band at 365 nm. These two bands are centered at 439 nm, corresponding to the position of III for $Mo_2$ carboxylates. Provided with the $\omega_{442}$ mode in the quantized scattering field (Figure 2C), we attribute these two bands to the polaritonic transitions to the dressed states P+ and P−, resulting from the resonant coupling of the $\delta \rightarrow \delta^*$ transition with light in $[Mo_2(TiPB)_4]^+$ (Figure 4B). Our interpretation of the anomalous absorption spectra for $Mo_2(TiPB)$ is strongly supported by the broad emission with a maximum at 550 nm for $Mo_2(O_2CCF_3)_4$ at low temperature (1.3 K), which exhibits a vibrational hyperfine structure of the Mo−Mo stretching frequency ($\omega_v \approx 400$ cm$^{-1}$).[54]

This long-lived emission does not energetically correspond to an electronic transition, nor does the broad absorption at 550 nm for $[Mo_2(TiPB)_4]^+$ and $[Mo_2(O_2CCH_3)_4]^+$ (**1**$^+$), according to the MO calculations.[41] The vibrational structure shown in the fluorescence transition clearly indicates that this 550 nm emission is due to vertical radiative decay of the excited state with an energy level significantly lower than the δ* state,[40,54,55] which is most likely the P− component of the Rabi doublet of the δ→δ* transition. On this basis, we located the P+ component of III for **1**$^+$ at 365 nm in the spectrum (Figure 4B) for resonance coupling to the $\omega_{442}$ mode of the scattering field, yielding a coupling strength of 9380 cm$^{-1}$, as observed for $[Mo_2(TiPB)_4]^+$. These results reveal coherent coupling between the δ→δ* transition and the scattering field in these carboxylate-supported complexes, demonstrating that the ligand environment plays a role in controlling the electron-photon interaction. Importantly, the optical coupling of the electronic transitions to the scattering field demonstrates that the quantum effects of the $Mo_2$ molecule as an individual quatum system can be manipulated by molecular design, showing the potential applications in quantum information processing.

**Resonant Coupling of the Mo−Mo Vibrational Mode with Photons Scattered by $Mo_2$.** Illumination of a single crystal **1** with a 532 nm laser beam produces an intense single scattering peak at 410 cm$^{-1}$ in the Raman spectrum (Figure 5A-a), signaling the Mo−Mo stretching mode ($\omega_v$).[40,43,54,55] For crystal **2**, multiple scattering peaks are found symmetrically distributed around the position of $\omega_v$ (Figure 5A-b). The intense peak at 407 cm$^{-1}$ is split at the tip (see Supplementary Figure 5), indicating the hyperfine structures of $\omega_v$ due to the presence of the nearly equivalent spin-1/2 states.[2] Ignoring this hyperfine structure, the crystal **2** exhibits a quintet structure with two pairs of sidebands centered at 407 cm$^{-1}$ (Figure 5A-c), indicating resonant coupling of the vibrational mode ($\omega_v$ = 407 cm$^{-1}$) to the scattered photons in the two-photon Raman transition (Figure 5C). We consider the first triplet, characterized by $\nu_{cent}$ at 407 cm$^{-1}$ and its satellites $\nu_{cent} \pm 26$ cm$^{-1}$, to be the primary Mollow triplet[20] due to the population decay of the dressed states of the photon number states $|1\rangle$ and $|2\rangle$ (Figure 5C), giving a vibrational coupling rate ($\Omega_v$) of 26 cm$^{-1}$ and $\eta = 0.03$ in the strong coupling regime.

The additional sidebands are displaced from $\nu_{cent}$ by 64 $cm^{-1}$, i.e., $\Omega = 64$ $cm^{-1}$, resulting from the high-level Fock states $|n\rangle$ with $n$ = 5 according to $\Omega_{5v} = \Omega_v\sqrt{n+1}$ [3] (Figure 5C, see Supplementary Table 4). Note that both sideband pairs are asymmetric with unequal integrated intensities. For the first pair, the $\nu_{blue}$ peak is more intense, while the $\nu_{red}$ peak has a low intensity. For the second pair, this relative intensity of the $\nu_{blue}$ vs. $\nu_{red}$ peaks is reversed. This gives the $\nu_{blue}$ and $\nu_{red}$ peaks similar total integrated intensities, and thus, the two triplets show comparable overall integrated sideband intensities. On this basis, we explain that the asymmetry of the sideband intensity is caused by the undesired population decay between the two scattering domains, corresponding to the Fock states $|1\rangle$-$|2\rangle$ and $|4\rangle$-$|5\rangle$ (Figure 5C).

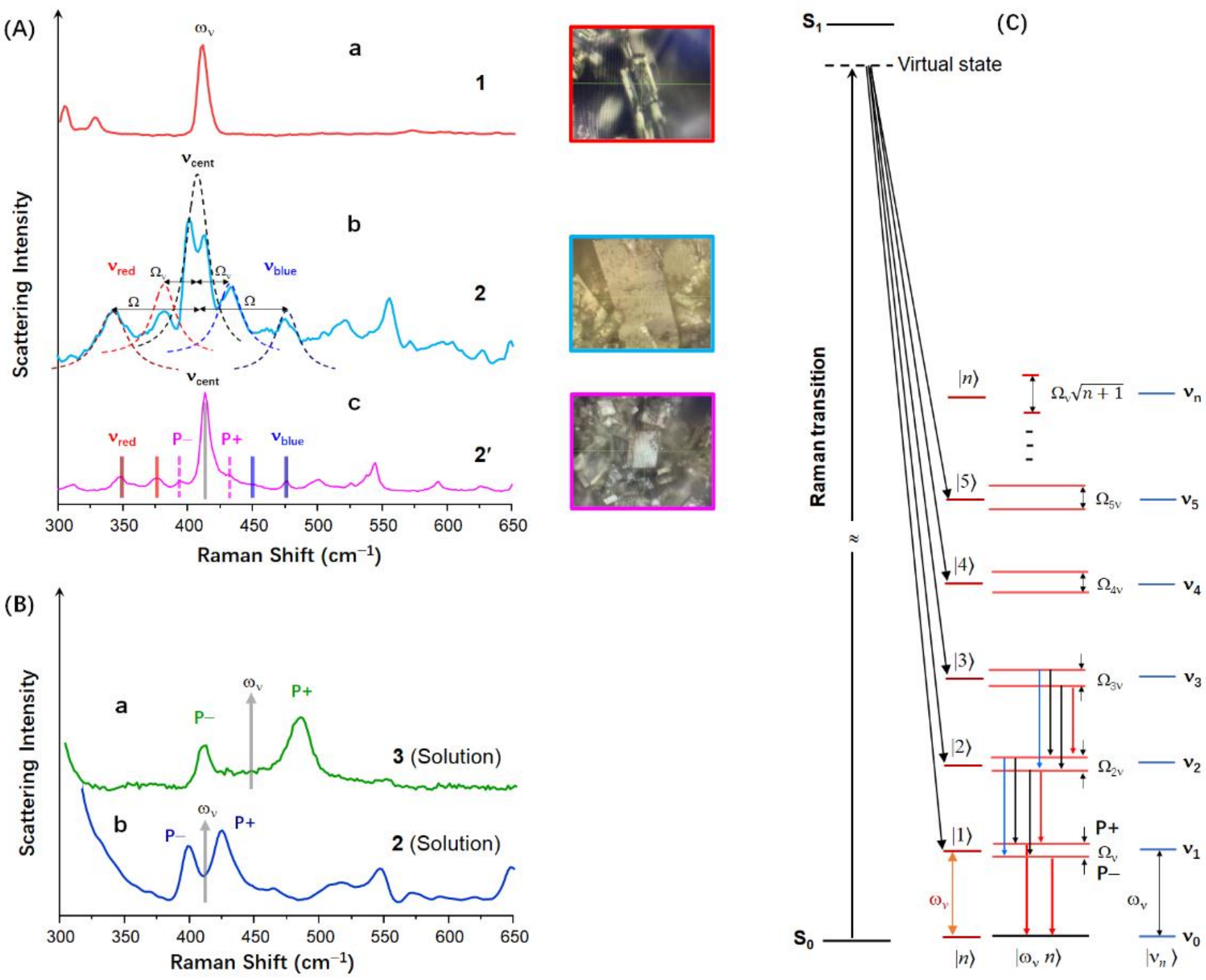

**Figure 5. Strong vibrational coupling for the $Mo_2$ complexes.** (A) Raman spectra of crystalline samples **1** (a), **2** (b) and **2′** (c). The measurements were performed by focusing a 532 nm laser beam on a single crystal, as shown in the insets. (B) Raman spectra of **2** (b) and **3** (a) in dichloromethane solution. (C) Schematic illustration of the formation of the Rabi doublet and the Mollow triplets of the Mo-Mo stretching mode for **2′** in the Raman scattering process.

In the Raman spectrum of the single crystal **2′** (Figure 5A-c), the intense peak at 414 $cm^{-1}$ represents the Mo-Mo stretching frequency ($\omega_v$), which is higher in energy than that of **2**. There are three pairs of side peaks symmetrically surrounding $\omega_v$. The separation between the two peaks of the first pair is similar to the distances of the second-pair peaks from the central peak, i.e., ~ 38 $cm^{-1}$. This indicates that the two peaks of the first pair belong to the Rabi branches P+ and P−, resulting from vibration-field coupling,[57] while the peaks of second-pair are assinged to the sidebands of the Mollow triplet of $\omega_v$, resulting in a similar coupling constant, i.e., $\Omega_v \approx 38$ $cm^{-1}$. This coupling constant is larger than that for **2** (26 $cm^{-1}$) presumably due to the higher frequency of Mo-Mo stretching, i.e., 414 $cm^{-1}$ vs. 407 $cm^{-1}$. The two peaks of the third pair are shifted from $\omega_v$ by 65 $cm^{-1}$ (Figure 5A-c, and see Supplementary Table 4), giving $\Omega = 65$ $cm^{-1}$ for the high-level Mollow triplet, characterized by $\omega_v$ and $\omega_v \pm \Omega$. This coupling strength is in excellent agreement with the prediction from $\Omega_{3v} = \Omega_v\sqrt{n+1}$ ($n = 2$),[3] indicating the second Mollow triplet resulting from coupling of $\omega_v$ to the Fock state $|2\rangle$ (Figure 5C and see Supplementary Table 4). Therefore, the first and second Mollow triplets are attributed to the scattering of the dressed state manifolds involving two adjacent Fock states $|1\rangle$-$|2\rangle$ and $|2\rangle$-$|3\rangle$ (Figure 5C), respectively, with relatively low population of the high-level states as indicated by the spectral asymmetry. To the best of our knowledge, up to date, a vibration-polariton Mollow triplet has not been observed experimentally, although vacuum Rabi splitting of a vibrational mode by zero EM filed has recently been demonstrated.[57,58] Significantly, the $Mo_2$ systems **2** and **2′**, through their characteristic Mo-Mo stretching modes, allow observation of the Jaynes-Cummings ladder structure for the dressed vibrational states.[9]

For **2** in solution, where the molecules are randomly oriented and freely moving, the Raman spectrum (Figure 5B-b) shows two isolated peaks at 400 $cm^{-1}$ and 425 $cm^{-1}$ centered at 412.5 $cm^{-1}$. These two peaks are assigned to the vibration-polaritons P+ and P− resulting from the vacuum Rabi splitting of $\omega_v$[58] by a single photon in the two-photon Raman transition (Figure 5C). The resonant coupling of $\omega_v$ to the scattered photon gives rise to a vibrational coupling rate $\Omega_v = 25$ $cm^{-1}$, similar to the coupling constant for the crystal **2** (Table S4). The vibrational coupling properties of **2** are further confirmed by the Raman scattering of the powder sample, which shows the combined scattering features of the single crystal and the solution (see Supplementary Figure 6). Complex **3** in solution exhibits the Rabi doublet composed of the P+ at 410 $cm^{-1}$ and the P− at 484 $cm^{-1}$ (Figure 5B-a), indicating a $\omega_v$ of 447 $cm^{-1}$. The large blue shift of $\omega_v$ and the increased $\Omega_v$ (74 $cm^{-1}$), compared to that of **1**, **2** and **2′**, can be attributed to the coordination of the electron-withdrawing carboxylate group in **3**. The observation of the Mollow triplet and the Rabi doublet of the Mo−Mo stretching mode indicates the strong resonant coupling of the vibrational state with the scattering field, strongly enhanced by the atomistic confinement,[13,24] analogous to the optical coupling of the electronic transitions observed in this $Mo_2$ molecular system.[27] The distinctive scattering spectrum profiles for the complexes in crystal and solution, which yield similar coupling constants, suggest that in both cases, the vibrational coupling occurs at the single molecule level. These results, tegether with the observed excitation-field coupling, reflecte the nature of the $M_2$ systems (M = Ni, Mo) as a Jaynes-Cummings molecule.[13]

**Discussion and Conclusion**

Sixty years after the discovery of quadruple bonds between metal atoms by F. A. Cotton, we propose in this work, based on the spectroscopic analysis, that the quadruply bonded $Mo_2$ unit of the complex molecules plays the role of a built-in cavity for field confinement. This study suggests that a single $Mo_2$ complex molecule can trap visible light photons between the quadruply bonded Mo atoms, creating an extremely strong local scattering field of quantum light across a broad spectrum of discrete single modes. Our findings indicate the $Mo_2$ unit as an individual quantum qubit, in which electronic, nuclear, and photonic degrees of freedom collectively define its optical behavior under ambient conditions. Spectral analysis reveals that the quantized scattering field of the $Mo_2$ unit drives the electronic and vibrational coupling of the molecule with light, profoundly modifying the electronic and vibrational structures. In this case, the involvement of photonic modes transforms the $Mo_2$ molecule into a many-body system, where traditional concepts of metal-metal bonding interaction and bond order are no longer exclusive descriptors of its behavior. The theoretical definition of bond order, which deals with the ground and excited electronic states of the dimetal center, may fail to fully capture the hybrid nature of this system, and cannot explain the discrepancies between experimental observations and theoretical results, nor the unpredictability of the spectroscopic results. Our study shows that, within these light-matter hybrid systems, the distinction between bonding and non-bonding interactions loses relevance. Instead, the energetics and spectroscopies of the hybrid system provide the most reliable confirmation of its properties. This insight is reinforced by the identical quantum optical behavior exhibited by both $Ni_2$ and $Mo_2$ system with distinct bonding characteristics.

Looking ahead, future work should focus on developing refined experimental techniques and theoretical methods for measuring and analyzing the scattering fields to gain further insight into the quantum states involved. This will help to validate the current findings and speculations, and explore applications in quantum technologies, such as quantum computing and communication. Such efforts will improve our understanding of light-matter interactions and advance both fundamental science and technological applications.

## Methods

### Chemical synthesis

The solvents used in this study were purified by the Inert solvent purification system, or by distilling over the drying reagents under inert atmosphere before used. All synthesis steps were carried out in Schleck line under nitrogen atmosphere. $Mo_2$ complexes $Mo_2(O_2CCH_3)_4$ (**1**),[59] $Mo_2(DAniF)_4$ (**2**)[42] (DAniF = *N*, *N′*-di(*p*-anisyl)formamidiante), $Mo_2(DAniF)_4^+$ (**2**$^+$), $Mo_2(DTolF)_4$ (**2′**) [42] (DTolF = *N*, *N′*-di(*p*-tolyl)formamidinate), $Mo_2(DTolF)_4^+$ (**2′**$^+$), $Mo_2(DAniF)_3(O_2CC_6H_5)$ (**3**) [60] and $\{Mo_2(DAniF)_3(O_2CC_6H_5)\}^+$ (**3**$^+$) were prepared by following the published methods.

### UV-visible absorption, fluorescence spectral measurements

The electronic (UV–Vis) spectra for the neutral and mono cationic $Mo_2$ complexes were recorded on Shimadzu UV-3600 (UV–VIS–NIR). All the spectroscopic measurements were conducted in dichloromethane (DCM) solution ($5 \times 10^{-3}$ mol $L^{-1}$) using quartz cell with a light path length of 2 mm. The fluorescence spectra of the complexes were recorded on Shimadzu RF-5301PC in DCM solution with 5 nm slit width and medium scan speed. $^{1}H$ NMR spectra were recorded on a Bruker-300 spectrometer.

**Electron paramagnetic resonance (EPR) chractrization**

The cationic $Mo_2$ complexes (0.5 mM in DCM) were prepared by oxidation of the neutral $Mo_2$ complexes. The samples were sealed in the 25–mm quartz tube for measurements (in situ) of electron paramagnetic resonance (EPR) spectra. The sepcta were recorded on a 9.7 GHz X–band Bruker BioSpin A300 spectrometer at 77K.

**Data availability**

All data are available in the manuscript or the supplementary materials. The spectra raw data have been deposited at Mendeley (Mendeley Data: https://doi.org/ 10.17632/mxbg5b62hp.1) and are publicly available as of the date of publication. Additional information about the data will be made available from the corresponding author upon reasonable request. No custom code or software was used in this study.

**Acknowledgments**

We acknowledge the primary financial support from the National Natural Science Foundation of China (22171107, 21971088, 21371074), Natural Science Foundation of Guangdong Province (2018A030313894), Jinan University, and the Fundamental Research Funds for the Central Universities.

**Author Contribution**

Chun Y. Liu conceived this project and designed the experiments and worked on the manuscript. Meng Miao carried out the major experimental work, including chemical synthesis, data collection and analysis, and prepared the Supplementary Information. Ying Ning Tan, Zi Cong He involved in the spectroscopic data analysis and assisted in manuscript preparation. Zi Hao Zhong, Yu Li Zhou, Jia Zhou, Guang Yuan Zhu were involved in experimental investigations.

## Competing interests

The authors declare no conflict of interest.

[56] Carmichael, H. J. & Walls, D. F. Proposal for the measurement of the resonant Stark effect by photon correlation techniques. J. Phys. B: Atom. Mol. Phys. 9, L43 (1976).

[57] Shalabney, A. Enhanced Raman Scattering from Vibro-Polariton Hybrid States. Angew. Chem. Int. Ed. 54, 7971–7975 (2015).

[58] Shalabney, A. et al. Coherent coupling of molecular resonators with a microcavity mode. Nat. Commun., 6, 5981 (2015).

[59] Cotton, F. A., Mester, Z. C. & Webb, T. R. Dimolybdenum tetraacetate. Acta Crystallogr B, 30, 2768–2770 (1974).

[60] Xiao, X. et al. Control of the charge distribution and modulation of the class II–III transition in weakly coupled Mo2–Mo2 systems. Inorg. Chem. 52, 12624–12633 (2013).